\documentclass[prd,preprint,tightenlines,superscriptaddress,floatfix,showpacs,preprintnumbers,nofootinbib,eqsecnum]{revtex4}

 \usepackage[dvips,final]{graphicx}
     \usepackage{epsfig}
      \usepackage{bm}

\begin{document}

\thispagestyle{empty} \preprint{\hbox{}} \vspace*{-10mm}

\title{Exclusive production of large invariant mass pion pairs 
in ultraperipheral ultrarelativistic heavy ion collisions}

\author{M.~K{\l}usek-Gawenda}
\email{mariola.klusek@ifj.edu.pl}

\affiliation{Institute of Nuclear Physics PAN, PL-31-342 Cracow, Poland}

\author{A.~Szczurek}
\email{antoni.szczurek@ifj.edu.pl}

\affiliation{Institute of Nuclear Physics PAN, PL-31-342 Cracow, Poland}
\affiliation{University of Rzesz\'ow, PL-35-959 Rzesz\'ow, Poland}

\date{\today}

\begin{abstract}

The cross section for exclusive production of $\pi^+ \pi^-$
and $\pi^0 \pi^0$ meson pairs in ultrarelativistic heavy ion collisions
is calculated for LHC energy $\sqrt{s_{NN}} =$ 3.5 TeV taking into account 
photon-photon mechanism. We concentrate on the production of large two-pion 
invariant masses where the mechanism of the elementary 
$\gamma \gamma \to \pi \pi$ process is not fully understood.
In order to include a size of nuclei we perform calculation
in the impact-parameter equivalent photon approximation (EPA).
Realistic charge densities are used to calculate charged form factor 
of $^{208}$Pb nucleus and to generate photon fluxes associated 
with ultrarelativistic heavy ions. 
Sizeable cross sections are obtained that can be measured at LHC.
The cross section for elementary $\gamma \gamma \to \pi \pi$ 
is calculated in the framework of pQCD Brodsky-Lepage (BL) mechanism
with the distribution amplitude used to descibe recent data
of the BABAR collaboration on pion transition form factor,
using hand-bag mechanism advocated to describe recent Belle
data as well as $t$ and $u$-channel meson/reggeon exchanges. 
We present distributions in two-pion invariant mass 
as well as the pion pair rapidity for the nuclear process.

\end{abstract}

\pacs{12.38.Bx, 
				 24.85.+p, 
                 25.20.Lj, 
                 25.75.Dw. 
	}

\maketitle

\section{Introduction}

It was shown in several review articles \cite{review}
that the ultrarelativistic collisions of heavy ions provide
a nice opportunity to study photon-photon collisions.
This is due to the enhancement caused by the large charge of the
colliding ions. Parametrically the cross section is proportional
to $Z_1^2 Z_2^2$ which is a huge number. It was discussed recently
that the inclusion of nuclei sizes as well as realistic charge distributions
in nuclei lowers the cross section compared to the naive predictions.
Recently we have studied the production of $\rho^0 \rho^0$ 
pairs \cite{KS_rho}, of muonic pairs \cite{KS_muon}, of heavy-quark 
heavy-antiquark \cite{KS_quark} as well as $D \bar{D}$
meson pair \cite{LS2010}.

In the present paper we wish to study probably the simplest to measure
exclusive production of pionic pairs. The elementary processes
$\gamma \gamma \to \pi^+ \pi^-$ and $\gamma \gamma \to \pi^0 \pi^0$
have been studied in detail in the past (see e.g. \cite{AS}).
While very low energies are the domain of the chiral perturbation
theory \cite{chpt}, at the intermediate energies one has to include also
pionic resonances in the s-channel as well $t$ and $u$-channel exchanges 
\cite{resonances, AS, ASinne, MNW}. 
At low dipion invariant masses a huge contribution
could come from a competitive photon--pomeron (pomeron--photon)
mechanism of exclusive $\rho^0$ production and it subsequent decay.
The cross section for this process is very large (see e.g. \cite{GM}). 
At even higher energies $\sqrt{s} >$ 2 GeV the mechanism 
of the reaction is not fully understood.
Brodsky and Lepage made a first prediction of the leading-order
pQCD \cite{BL81} which was further studied e.g. in \cite{JA,Nizic}.
In general, the predictions of the pQCD calculation lay below the
experimental data measured at LEP \cite{Aleph} and recently by the
Belle collaboration \cite{Belle}. The next-to-leading order calculation 
has been carried out only in Ref. \cite{NLOpQCD} and their result 
is not able to describe the present experimental data. 
The pQCD amplitude for the $\gamma \gamma \to \pi \pi$ reaction 
depends on the pion distribution amplitude. It was believed 
for already some time that the pion distribution
amplitude is close to the asymptotic form (6 $x (1-x)$).
This turned out to be inconsistent with recent results of the
BABAR collaboration for the pion transition form factor 
$F_{\gamma^* \gamma \pi}$ for large photon virtualities \cite{BABAR}.
The authors of Ref. \cite{WH} used a new model of the distribution
amplitude which can describe the BABAR data.
We shall use this model for the $\gamma \gamma \to \pi \pi$ reaction.

Some time ago Diehl, Kroll and Vogt (DKV) suggested that 
a soft hand-bag mechanism may be the dominant mechanism \cite{HB} 
for wide-angle scattering at intermediate energies. 
In this approach the normalization as well as energy dependence 
of the corresponding cross section are adjusted to the world-data on the 
$\gamma \gamma \to \pi^+ \pi^-$ production \cite{HB}.

In the present paper first we show how the different mechanisms
describe the elementary data. Next we present our predictions for
the nucleus-nucleus collisions. We will show distributions in the
dipion invariant mass as well as in the pion pair rapidity.
These are quantities which can be easily calculated in the 
impact-parameter equivalent photon approximation (b-space EPA).

\section{Elementary cross section for $\gamma \gamma \to \pi \pi$}
\subsection{Perturbative QCD approach}
\begin{figure}[!h]                   
\begin{minipage}[t]{0.3\textwidth}
\centering
\includegraphics[width=1\textwidth]{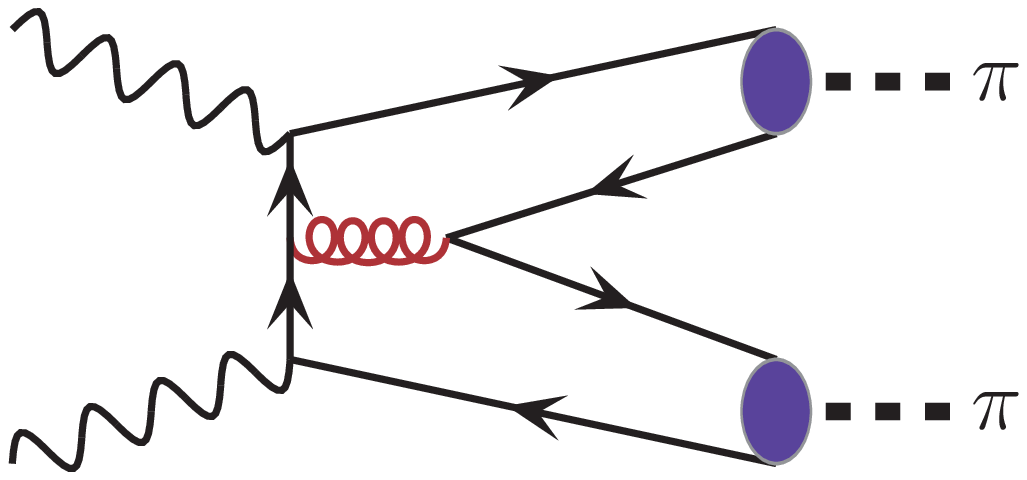}
\end{minipage}
\hspace{0.03\textwidth}
\begin{minipage}[t]{0.3\textwidth}
\centering
\includegraphics[width=1\textwidth]{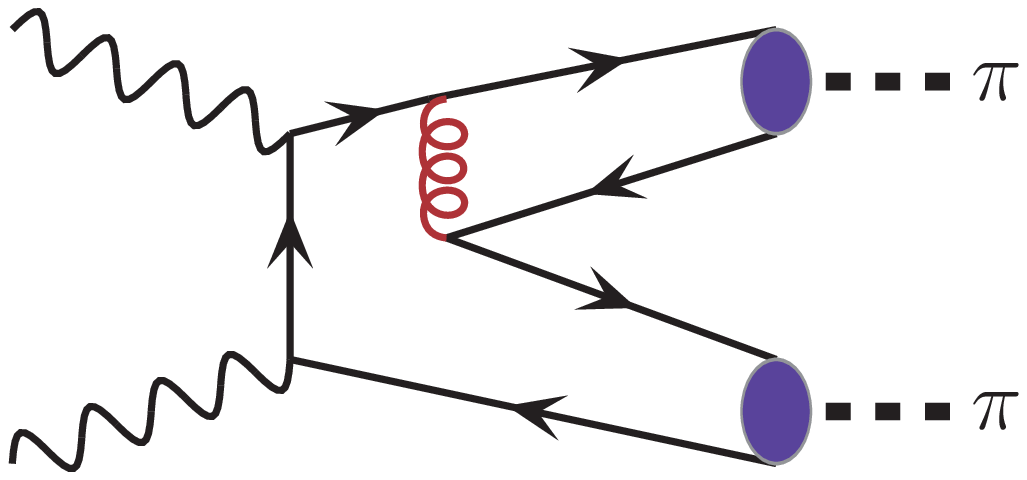}
\end{minipage}
\hspace{0.03\textwidth}
\begin{minipage}[t]{0.3\textwidth}
\centering
\includegraphics[width=1\textwidth]{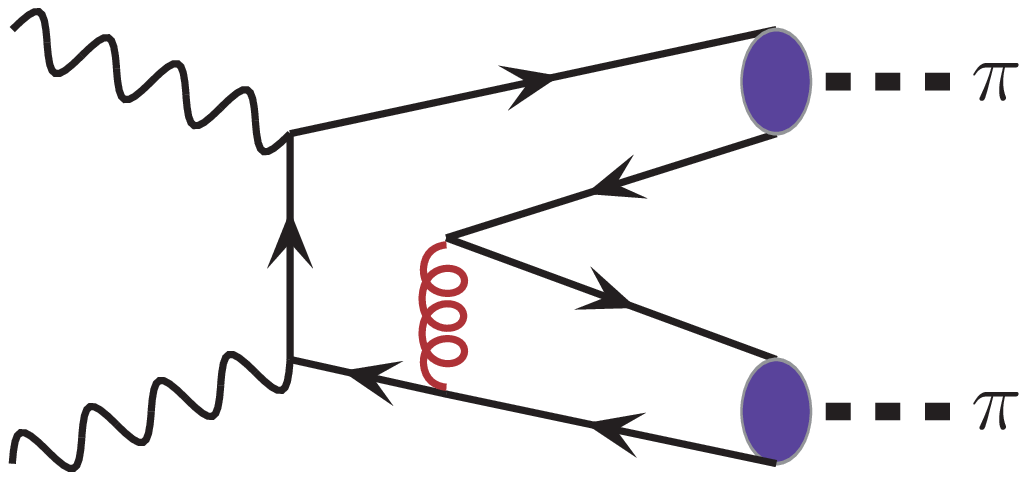}
\end{minipage}

\begin{minipage}[t]{0.3\textwidth}
\centering
\includegraphics[width=1\textwidth]{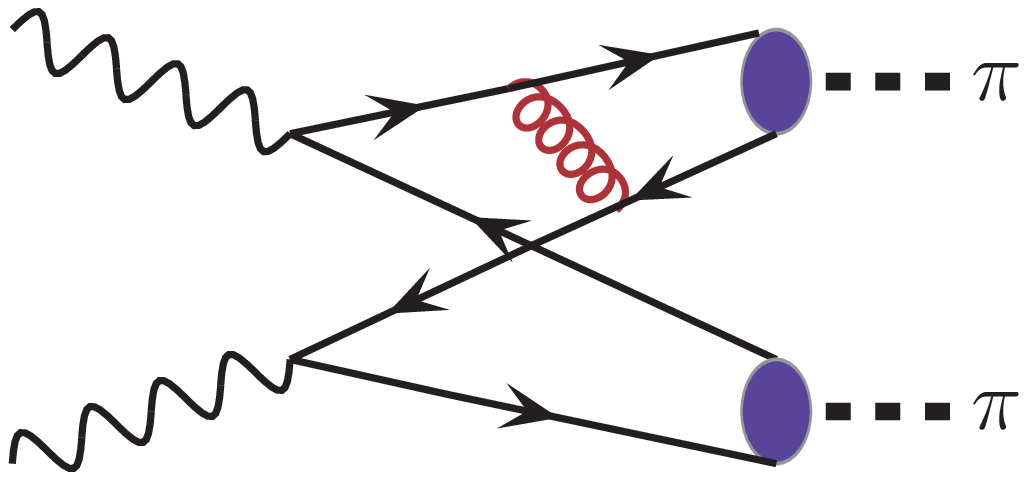}
\end{minipage}
\hspace{0.03\textwidth}
\begin{minipage}[t]{0.3\textwidth}
\centering
\includegraphics[width=1\textwidth]{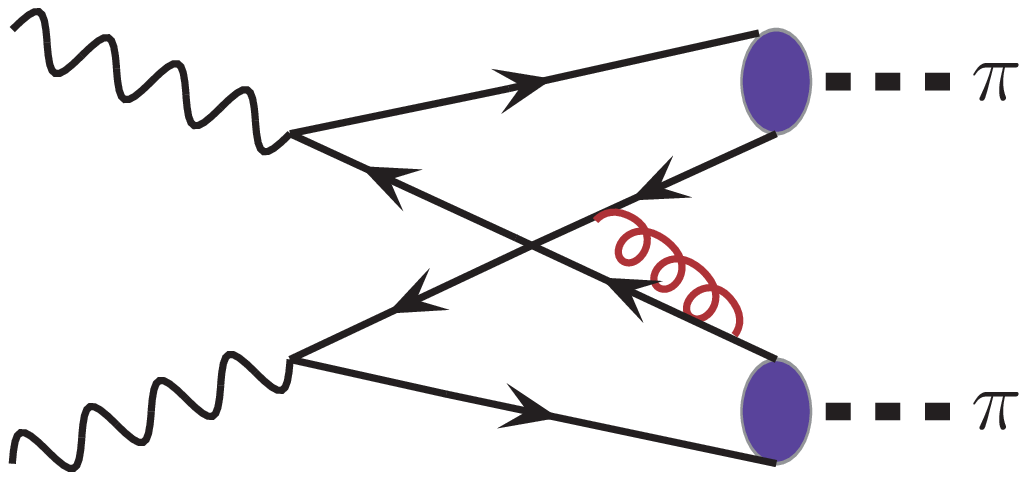}
\end{minipage}
   \caption{\label{fig:Feynman}\textsl{}
   \small Feynman diagrams describing 
   the $\gamma \gamma \to ( q \bar{q}) (q \bar{q}) \to \pi \pi$ 
   amplitude in the LO pQCD.
}
\end{figure}
Basic diagrams of the Brodsky and Lepage formalism are shown 
in Fig. \ref{fig:Feynman}. The invariant amplitude for the initial 
helicities of two photons can be written as the following convolution:
\begin{eqnarray}
 \mathcal{M}  \left( \lambda_1, \lambda_2 \right)  = 
 \int_0^1 dx \int_0^1 dy \, \phi_\pi \left( x, \mu^2_x \right) 
 T^{\lambda_1 \lambda_2}_H \left( x,y, \mu^2 \right) 
 \phi_\pi \left( y, \mu^2_y \right),
 \label{eq.amp}
\end{eqnarray}
where
$\mu_x = min \left( x, 1-x  \right) \sqrt{s(1-z^2)}$,
$\mu_y = min \left( y, 1-y  \right) \sqrt{s(1-z^2)}$; $z= \cos \theta$ \cite{BL81}.
We take the helicity dependent hard scattering amplitudes from Ref.~\cite{JA}.
These scattering amplitudes are different for $\pi^+ \pi^-$
and $\pi^0 \pi^0$.
It was proposed in Ref. \cite{SS2003} to exclude the region of small Mandelstam 
$t$ and $u$ variables by multiplying the pQCD amplitude (\ref{eq.amp}) 
by an extra form factor which cuts off the soft regions
which were taken into account in Ref. \cite{SS2003} explicity by including
meson exchanges.
The following form of the form factor was proposed in \cite{SS2003}:
\begin{equation}
 F^{pQCD}_{reg} \left(t,u \right)= 
 \left[ 1-\exp \left( \frac{t-t_m}{\Lambda_{reg}^2} \right) \right] 
 \left[ 1-\exp \left( \frac{u-u_m}{\Lambda_{reg}^2} \right)  \right],
 \label{eq.ff_pqcd}
\end{equation}
where $t_m=u_m$ are the maximal kinematically allowed values of $t$ and $u$. 
$\Lambda_{reg}$ is a cut-off parameter expected to be of the order of $1$ GeV. 
The distribution amplitudes are subjected to the ERBL pQCD evolution \cite{ER, BL79}.
The scale dependent quark distribution amplitude of the pion \cite{Muller,AB}
can be expanded in term of the Gegenbauer polynomials:
\begin{equation}
\phi_\pi \left( x, \mu^2 \right) = 
\frac{f_\pi}{2 \sqrt{3}}
6x \left( 1-x \right) {\sum_{n=0}^{\infty}} '  C_n^{3/2} 
\left(2x-1 \right) a_n \left(\mu^2 \right)  ,
\end{equation}
where the expansion coefficients (only even above) can be written as:
\begin{equation}
a_n \left(\mu^2 \right)  = 
\frac{2}{3} \frac{2n+3}{ \left(n+1 \right) \left( n+2 \right)} 
\left( \frac{\alpha \left(\mu^2 \right)}
{\alpha \left(\mu_0^2 \right)}\right)^{- \frac{C_F}{\beta_0} 
\left[ 3+ \frac{2}{\left(n+1 \right) \left(n+2 \right)} 
- 4 \sum_{k=1}^{n+1} \frac{1}{k} \right]} \int_0^1 dx C_n^{3/2} 
\left(2x-1 \right) \phi_\pi \left(x,\mu_0^2 \right), 
\end{equation}
where $\beta_0 = \frac{11}{3} C_A - \frac{2}{3} N_F$, 
$\alpha_s \left(\mu^2 \right) = \frac{4 \pi}{\beta_0 \ln \frac{\mu^2}{\Lambda^2_{QCD}}}$, 
$C_n^{3/2}$ denote the Gegenbauer polynomials, 
$C_F=\frac{4}{3}$, $C_A =$ 3, $N_F$ is the number of active quarks 
and $\Lambda$ is the QCD scale parameter. \\

Different distribution amplitudes have been used in the past 
\cite{BL81,Ch,AB}.
Wu and Huang \cite{WH} proposed recently a new distribution amplitude 
(based on a certain light-cone wave function):
\begin{eqnarray}
 \phi_\pi \left( x, \mu_0^2 \right) & = 
 & \frac{\sqrt{3}A \, m_q \beta}{2 \sqrt{2} \pi^{3/2} f_\pi} 
 \sqrt{x \left( 1-x \right)} 
 \left( 1+B \times C_2^{3/2} \left(2x-1 \right) \right) \nonumber \\
 &\times& \left( \mbox{Erf} \left[ \sqrt{\frac{m_q^2+\mu_0^2}
 {8 \beta^2 x \left(1-x \right)}} \right] 
 - \mbox{Erf} \left[ \sqrt{\frac{m_q^2}{8 \beta^2 x \left(1-x \right)}} 
 \right]\right).
 \label{eq.WH}
\end{eqnarray}
This pion distribution amplitude at the initial scale is controlled by 
the parameter B. It has been found that the BABAR data 
at low and high energy regions can be described by setting B to be around 0.6. 
This pion distribution amplitude is rather close to the well know 
Chernyak-Zhitnitsky \cite{CZ} distribution amplitude 
($\phi_{\pi \, CZ} = 30x(1-x)(2x-1)^2$).
In the following (Eq.~\ref{eq.WH}) we shall use $B=$ 0.6 and $m_q=$ 0.3 GeV. 
Then $A=$ 16.62 GeV$^{-1}$ and $\beta=$ 0.745 GeV.
$f_\pi$ above is the pion decay constant. 

The total (angle integrated) cross section 
for the process can be expressed in terms 
of the amplitude of the process discussed above as:
\begin{equation}
\sigma \left(\gamma \gamma \to \pi \pi \right) = 
\int \frac{2 \pi}{4 \cdot 64 \pi^2 W^2 } \frac{p}{q} 
\sum_{\lambda_1, \lambda_2} 
\left|  \mathcal{M}  \left( \lambda_1, \lambda_2 \right) \right|^2 dz \;,
\end{equation}
where the factor $4$ is due to averaging over initial photon helicities.
\subsection{Hand-bag model}

The hand-bag model was proposed as an alternative
for the leading term BL pQCD approach \cite{HB}.
It is based on the philosophy that the present
energies are not sufficient for the dominance 
of the leading pQCD terms. As in the case of 
BL pQCD the hand-bag approach applies at large
Mandelstam variables $s \sim -t \sim -u$ i. e. at large 
momentum transfers. Diehl, Kroll and Vogt presented 
a sketchy derivation \cite{HB} obtaining that
the angular dependence of the amplitude is 
$ \propto 1/ \sin^2 \theta$. Then the cross section integrated 
over $\cos \theta$ 
from $- \cos \theta_0$ to $\cos \theta_0$ 
for a charged pion pairs takes the simple form:
\begin{equation}
\sigma \left( \gamma \gamma \to \pi^+ \pi^- \right)  =
\frac{4 \pi \alpha_{em}^2}{s} \left( \frac{\cos \theta_0}{\sin^2 \theta_0} 
+ \frac{1}{2} \ln \frac{1+ \cos \theta_0}{1- \cos \theta_0} \right) 
\left| R_{\pi \pi} \left(s \right) \right|^2.
\end{equation}
Additionally, the ratio of the cross section for the $\pi^0 \pi^0$
process to the $\pi^+ \pi^-$ process doesn't depend on $\theta$ 
and is $\frac{1}{2}$. The nonperturbative object 
$R_{\pi \pi} \left(s \right)$ describing transition from a quark
pair to a meson pair cannot be calulated from 
first principles.
In Ref. \cite{HB} the form factor
was parametrized in terms of the valence 
and non-valence form factors as:
\begin{equation}
R_{\pi \pi} \left(s \right) = 
\frac{5}{9s} a_u \left( \frac{s_0}{s} \right)^{n_u} + 
\frac{1}{9s} a_s \left( \frac{s_0}{s} \right)^{n_s},
\end{equation}
where the authors of \cite{HB} have chosen $s_0 = $ 9 GeV$^2$.
The $a_u$, $n_u$, $a_s$ and $n_s$ values 
found from the fit in Ref. \cite{HB}
slightly depend on energy.
For simplicity we have averaged these values 
and used in the present calculations:
$a_u=$ 1.375 GeV$^2$, $n_u=$ 0.4175, 
$a_s=$ 0.5025 GeV$^2$ and $n_s=$ 1.195.
The hand-bag approach was criticised in 
Ref. \cite{Ch}.
%
\subsection{Meson exchanges in t or u channels}

Since several mesons ($\rho$, $\omega$, $a_1$, $a_2$, $b_1$)
decay into $\gamma \pi$ channels this means that $t$ and/or $u$ channel
exchanges of their virtual (space-like) counterparts may be important
for the $\gamma \gamma \to \pi^+ \pi^-$ and $\gamma \gamma \to \pi^0 \pi^0$
reactions.
As an example in the following we consider $\omega$ exchange
for the $\pi^0 \pi^0$ channel. $\rho$ meson exchange also contributes
to this reaction
but its contribution is much lower (the corresponding coupling constant
is 3 times smaller than that for $\omega$ meson exchange
and it enters here in the second power already
in the amplitude). So far the exchange of the tensor mesons was not
discussed in detail in the literature.

The amplitude for the $\gamma \gamma \to \pi \pi$ reaction via
vector meson exchange can be calculated by means of standard
Feynman rules assuming tensorial form of the $V \pi \gamma$ coupling.
The corresponding coupling constant can be obtained by fitting 
$V \to \pi \gamma$ decay width.
A simple and compact formula for the omega meson exchange amplitude 
was presented e.g. in Ref. \cite{MNW}.
It can be written as:
\begin{equation}
\mathcal{M} \left( \lambda_1, \lambda_2 \right)  =
  \frac{\alpha_{em} h_{\omega}^2}{16} 
  \left( X_t \left( \lambda_1, \lambda_2 \right) + 
  X_u \left( \lambda_1, \lambda_2 \right) \right),
\end{equation}
\begin{equation}
X_t \left( \lambda_1, \lambda_2 \right) = 
\frac{\epsilon_1 \left( \lambda_1 \right) \epsilon_2 \left( \lambda_2 \right)
 \left\{t \left(s-u \right) + m_{\pi}^4   \right\}
-2s \left\{\epsilon_1 \left( \lambda_1 \right) p_1 \right\} 
\left\{\epsilon_2 \left( \lambda_2 \right) p_2 \right\}}
{t-m_{\omega}^2} F_{\omega}^2 \left(t \right),
\end{equation}
\begin{equation}
X_u \left( \lambda_1, \lambda_2 \right) = 
\frac{\epsilon_1 \left( \lambda_1 \right) \epsilon_2 \left( \lambda_2 \right) 
\left\{u \left(s-t \right) + m_{\pi}^4   \right\}
-2s \left\{\epsilon_1 \left( \lambda_1 \right) p_1 \right\} 
\left\{\epsilon_2 \left( \lambda_2 \right) p_2 \right\}}
{u-m_{\omega}^2} F_{\omega}^2 \left(u \right),
\end{equation}
where the size of the radiative coupling 
was obtained from the ratiative decay $\omega \to \pi^0 \gamma$.
In contrast to Ref. \cite{MNW} we include also vertex form factors
($F_{\omega}(t)$, $F_{\omega}(u)$) which take into account the extended nature 
of the particles involved off-shell effects as well as high-energy reggezation. 
Not including the form factors leads, in our opinion, to nonphysical results, 
especially at large energies. 
Above $W >$ 1.5 GeV the so-calculated cross section would significantly 
exceed experimental data.
This point was not disccused in the literature as previous analyses were limited to
rather low energies where the problem was not visible (note a remark in Ref. \cite{AS}).

Using a vector particle propagators at high energy is not sufficient
and one has to include reggezation.
This is included in our calculation by multiplying the $t$ and/or $u$-exchange amplitudes
by the extra energy dependent factors: 
\begin{equation}
F_{\omega}\left(t/u \right)=
\exp \left(\frac{t/u-m_{\omega}^2}{2 \Lambda_{\omega}^2} \right) 
\left(\frac{s}{s_0} \right)^{\alpha \left(t/u \right)-1}.
\label{eq.ffomega}
\end{equation}
The $\omega$ trajectory is parametrized as 
$\alpha(t/u)=0.64+0.8 \, t/u$ \cite{SHHKM} 
and $s_0=$ 1 GeV is taken in further calculations.


\subsection{Results}
\begin{figure}[!h]               
\begin{minipage}[t]{0.46\textwidth}
\centering
\includegraphics[width=0.8\textwidth]{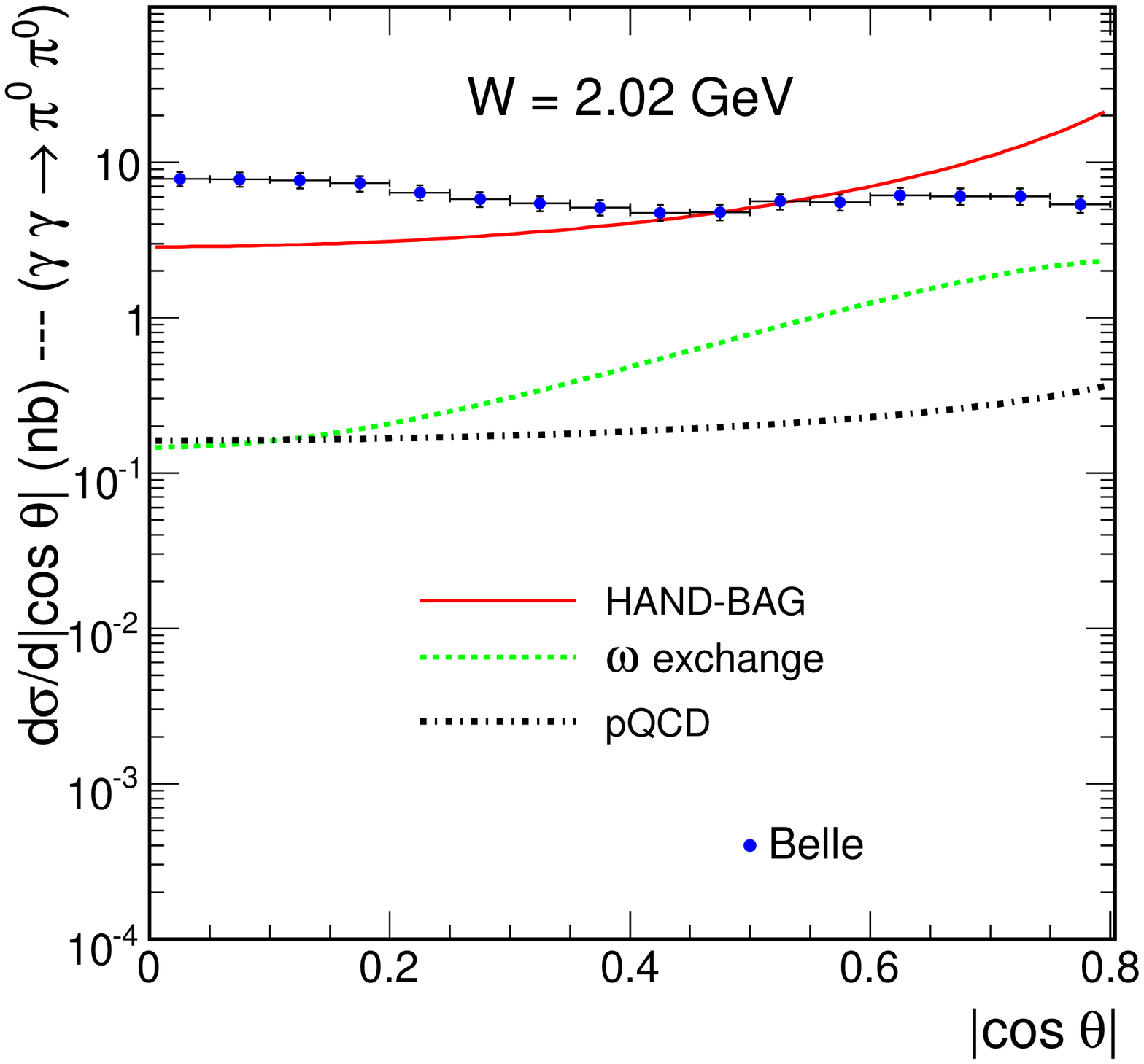}
\end{minipage}
\hspace{0.03\textwidth}
\begin{minipage}[t]{0.46\textwidth}
\centering
\includegraphics[width=0.8\textwidth]{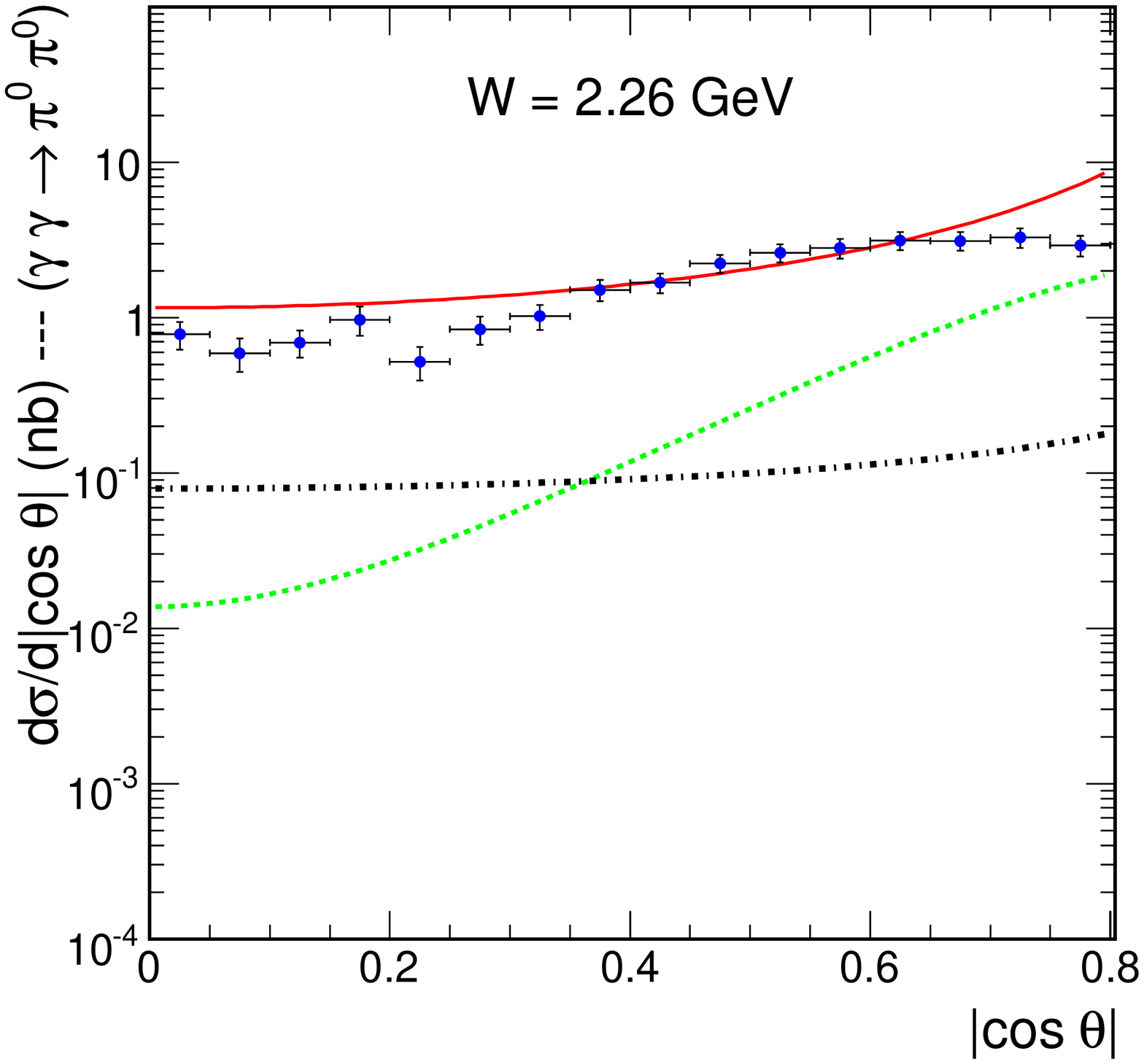}
\end{minipage}
\begin{minipage}[t]{0.46\textwidth}
\centering
\includegraphics[width=0.8\textwidth]{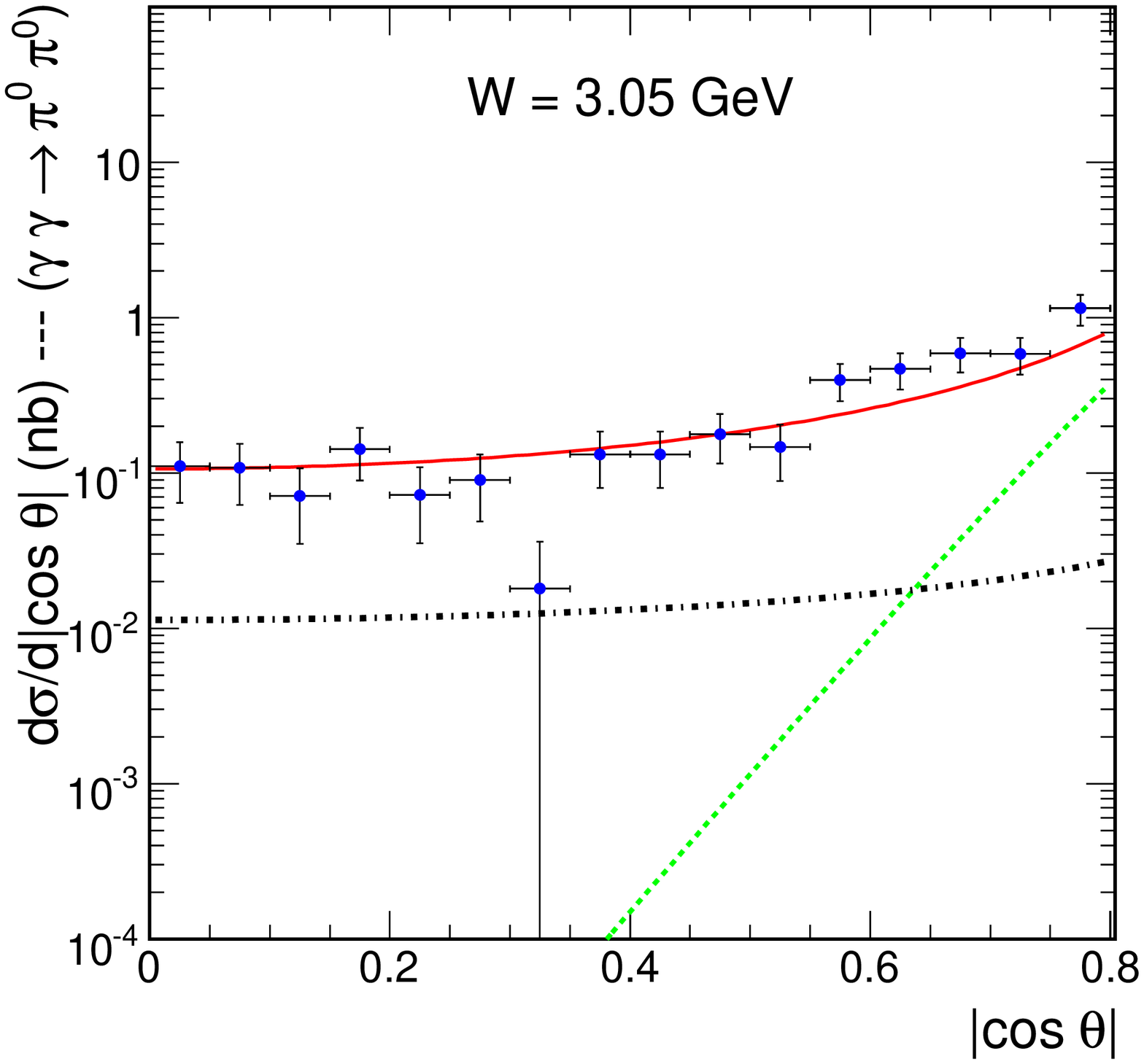}
\end{minipage}
\hspace{0.03\textwidth}
\begin{minipage}[t]{0.46\textwidth}
\centering
\includegraphics[width=0.8\textwidth]{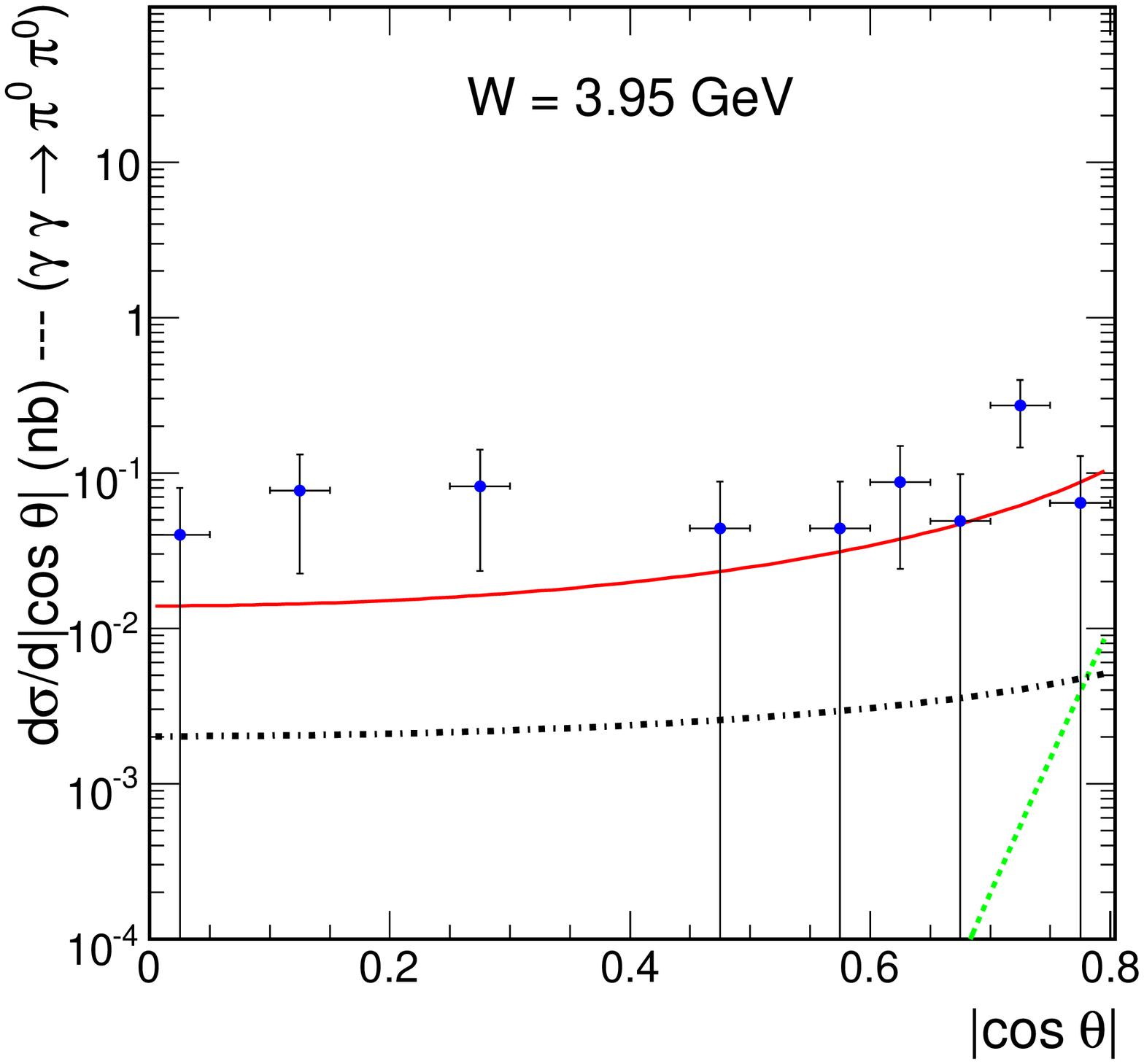}
\end{minipage}
   \caption{\label{fig:el_cross_section_dz}
   \small Angular distributions for the $\gamma \gamma \to \pi^0 \pi^0$
   reaction. 
   The experimental data are taken from Ref. \cite{Belle_dz}.}
\end{figure}
%
In Fig. \ref{fig:el_cross_section_dz} we show the predictions of the
hand-bag approach (solid lines), reggeized $\omega$ - exchange (dotted lines) 
and the Brodsky - Lepage pQCD approach (dashed lines) for angular distributions of the
$\gamma \gamma \to \pi^0 \pi^0$ reaction for $W$ = 2.02, 2.26, 3,05 3.95 GeV.
The pQCD results have been calculated in the case when $F_{reg}^{pQCD}=$ 1.
The cut-off parameter $\Lambda_{\omega}$ in Eq. (\ref{eq.ffomega})
was taken to be $\Lambda_{\omega}=$ 1 GeV.
The results of different calculation are confronted with the Belle data.
For the energies of present experiments the pQCD result is well below 
the experimental data. As can be seen from the figure
the $\omega$ - exchange may play a role 
only at large $\vert \cos \theta \vert$.
The result of the hand-bag approach starts to describe the data at energies
$\sqrt{s} >$ 3 GeV.

\begin{figure}[!h]               
\begin{minipage}[t]{0.46\textwidth}
\centering
\includegraphics[width=1\textwidth]{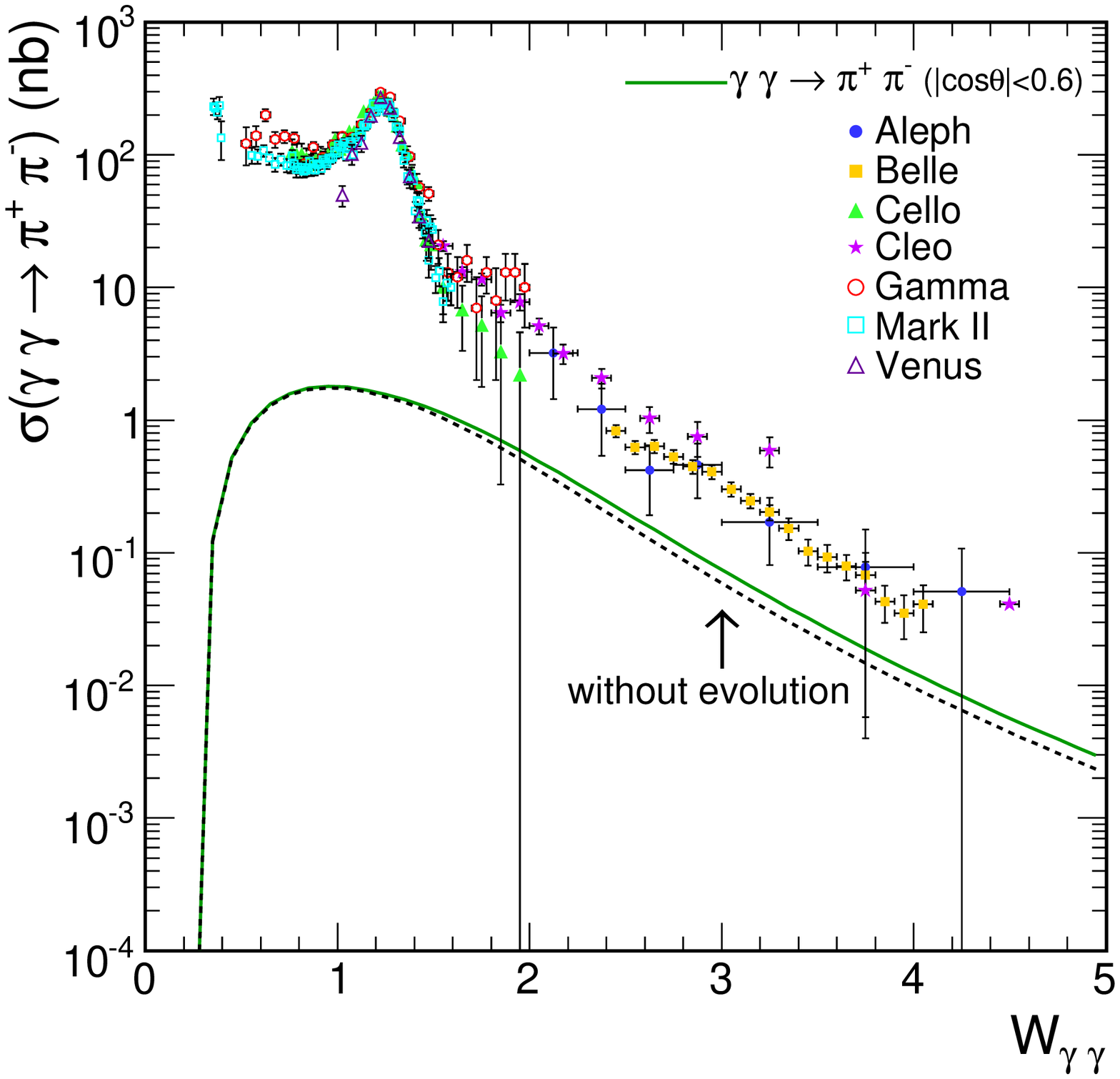}
\end{minipage}
\hspace{0.03\textwidth}
\begin{minipage}[t]{0.46\textwidth}
\centering
\includegraphics[width=1\textwidth]{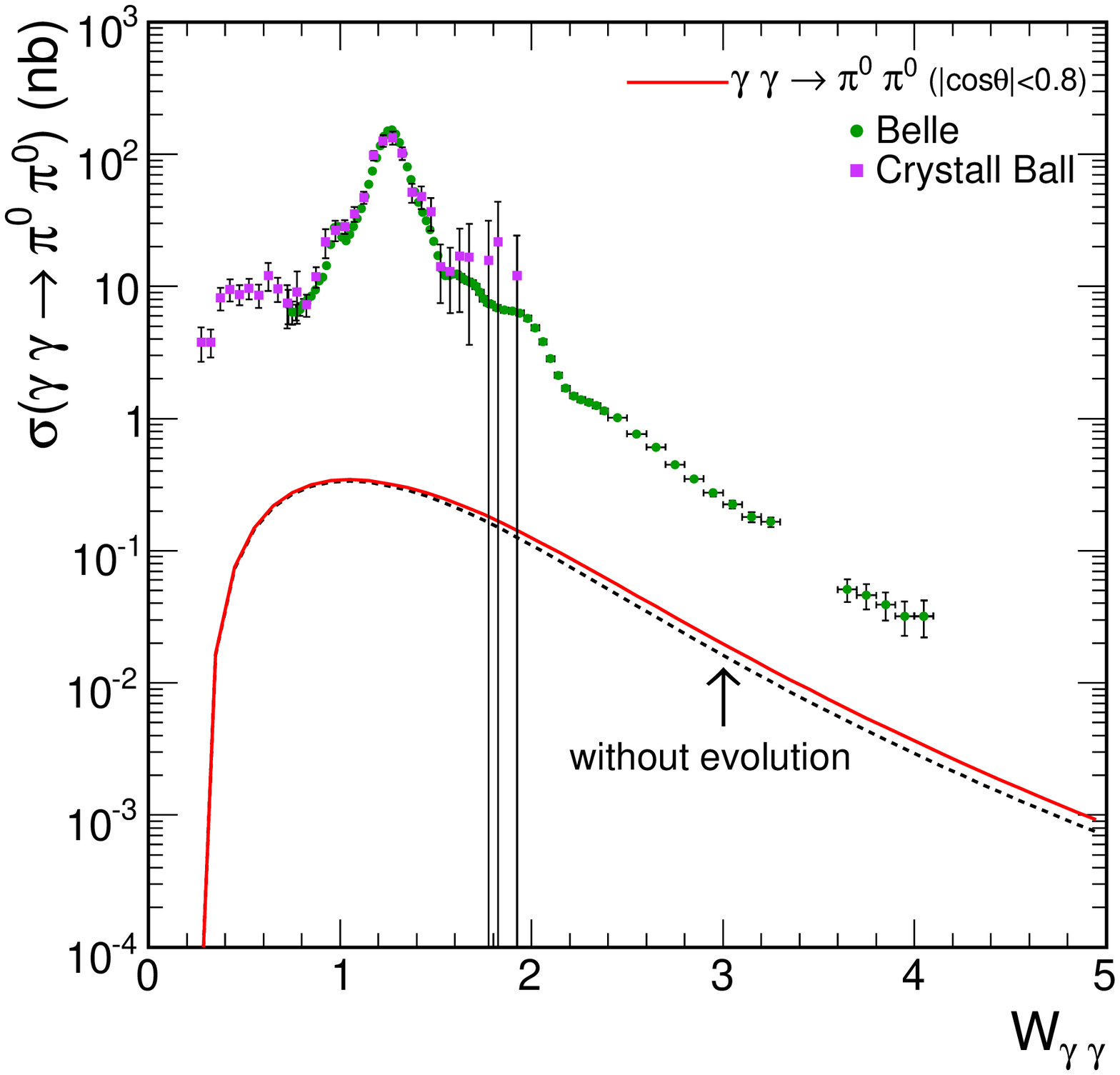}
\end{minipage}
   \caption{\label{fig:el_cross_section}
   \small The pQCD cross section for $\gamma \gamma \to \pi^+ \pi^-$ (left panel)
   and for $\gamma \gamma \to \pi^0 \pi^0$ (right panel)
   as a function of photon-photon energy.
   The solid lines show the results for evolved $\phi_\pi \left( x, \mu^2 \right)$
   and the dashed lines are for $\phi_\pi \left( x, \mu_0^2 \right)$
   where $\mu_0^2 = $ 0.25 GeV$^2$ was chosen. }
\end{figure}
%
In Fig. \ref{fig:el_cross_section} we compare
the pQCD $\gamma \gamma \to \pi \pi$ cross section 
for the pion distribution amplitude with and without pQCD evolution.
The effect of the pQCD evolution on the angle-integrated 
cross section is very small, practically negligible.
The data correspond to limited angular ranges 
given in the figure.
The data for the $\gamma \gamma \to \pi^+ \pi^-$ reaction are from
the ALEPH \cite{Aleph}, Belle \cite{Belle}, CELLO \cite{Cello},
CLEO \cite{Cleo}, Gamma \cite{Gamma}, Mark II \cite{Mark}
and VENUS \cite{Venus} Collaborations. 
For the $\gamma \gamma \to \pi^0 \pi^0$ reaction
we present the Belle \cite{Belle_08} and Crystall Ball \cite{CBall} data.

\begin{figure}[!h]             
\begin{minipage}[t]{0.46\textwidth}
\centering
\includegraphics[width=1\textwidth]{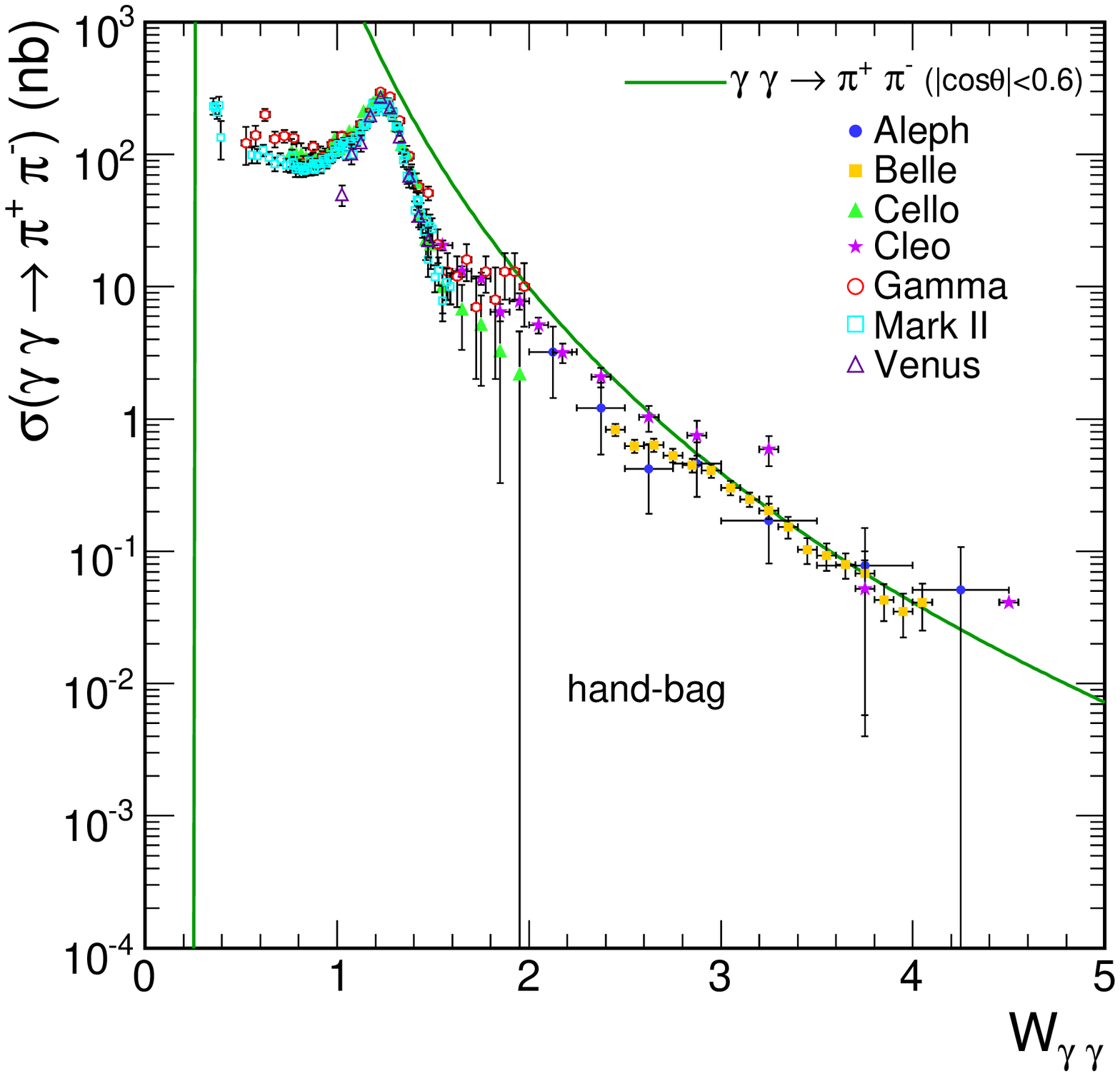}
\end{minipage}
\hspace{0.03\textwidth}
\begin{minipage}[t]{0.46\textwidth}
\centering
\includegraphics[width=1\textwidth]{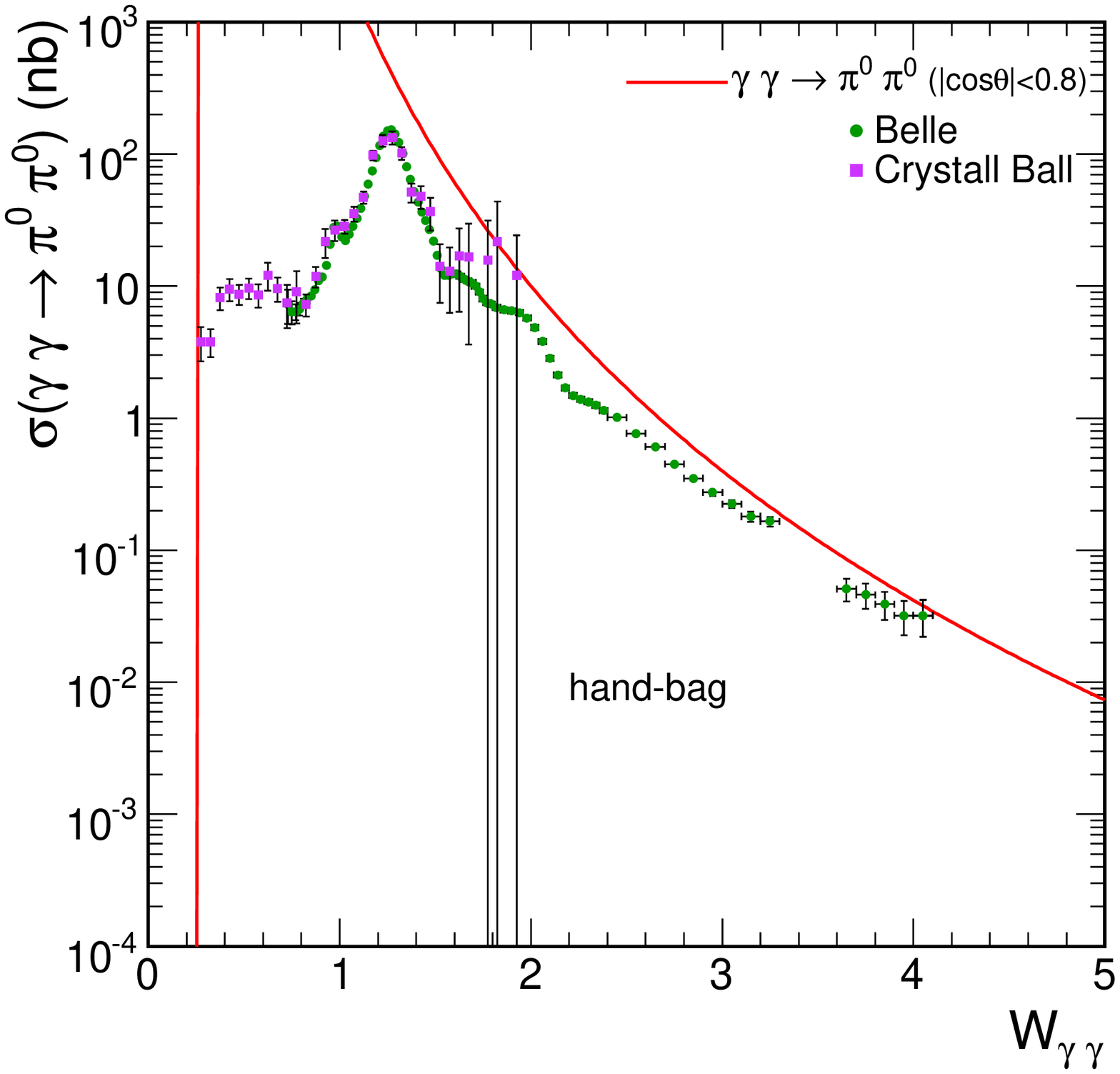}
\end{minipage}
   \caption{\label{fig:el_cross_section_hb}
   \small The hand-bag contribution for $\gamma \gamma \to \pi^+ \pi^-$ (left panel)
   and for $\gamma \gamma \to \pi^0 \pi^0$ (right panel)
   as a function of photon-photon energy.
           }
\end{figure}
%
In Fig. \ref{fig:el_cross_section_hb} we show the predictions of
the hand-bag approach \cite{HB} together with modern experimental
data. The predictions can be taken seriously above the resonance
region, i.e. when $\sqrt{s_{\gamma \gamma}} >$ 2.5 GeV. The parameters of the hand-bag 
contribution were adjusted to somewhat older experimental data.
One can see that the hand-bag approach, while consistent
with the $\pi^+ \pi^-$ data, slightly overestimates
the $\pi^0 \pi^0$ data.

\begin{figure}[!h]             
\begin{minipage}[t]{0.46\textwidth}
\centering
\includegraphics[width=1\textwidth]{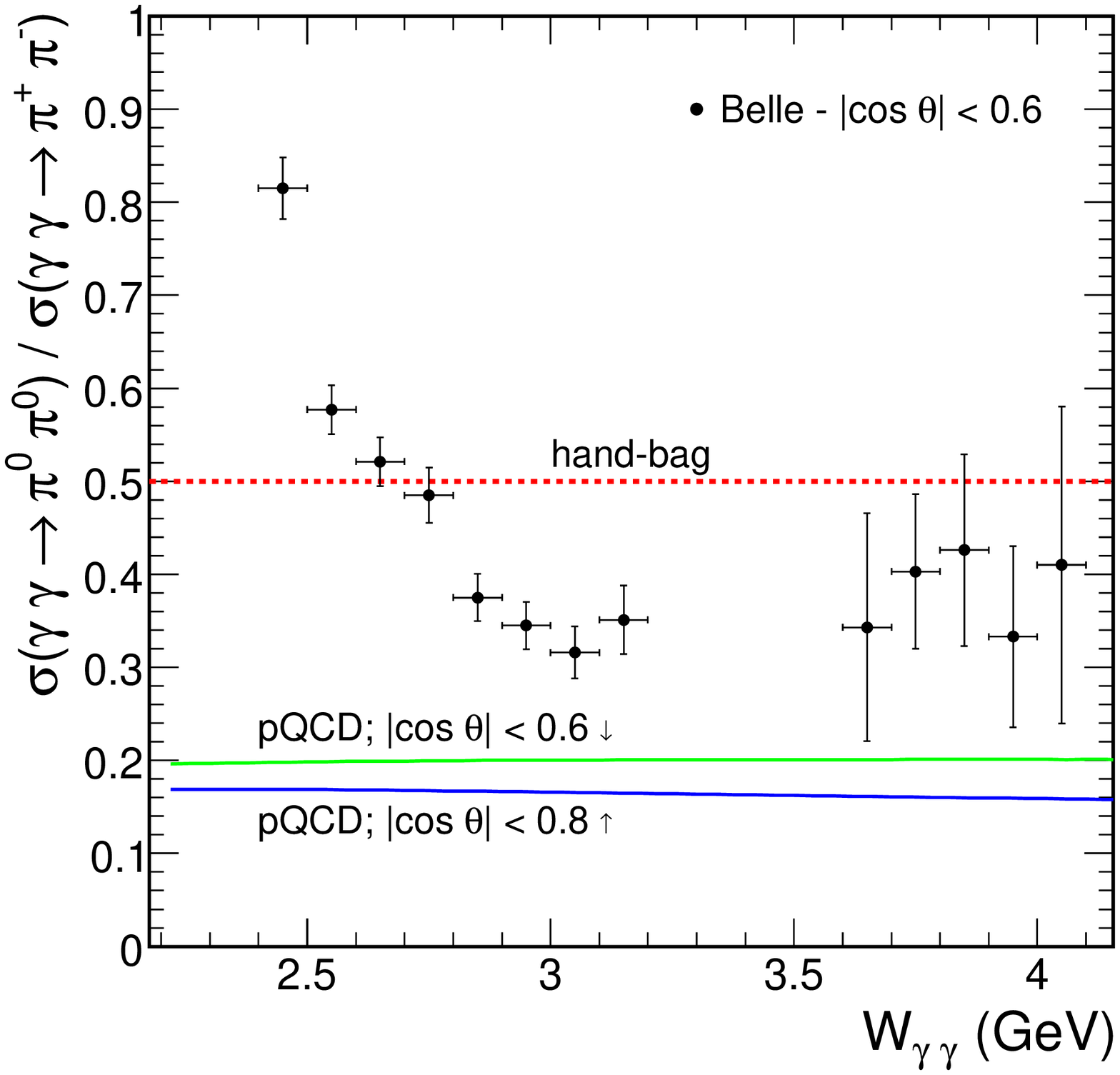}
\end{minipage}
   \caption{\label{fig:ratio_elementary}
   \small Ratio of the cross section for the 
   $\gamma \gamma \to \pi^0 \pi^0$ process to that for the
   $\gamma \gamma \to \pi^+ \pi^-$ process.
   The experimental data were obtained based on the
   original Belle Colaboration \cite{Belle,Belle_08} data
   as explained in the text.
           }
\end{figure}
%
In Fig. \ref{fig:ratio_elementary} we show the ratio of the
cross section for the $\gamma \gamma \to \pi^0 \pi^0$ 
process to that for the $\gamma \gamma \to \pi^+ \pi^-$ process.
The dashed line represents the hand-bag model \cite{HB} result
and the solid lines is for the Brodsky-Lepage pQCD approach. 
For larger range of $z=\cos \theta$
the ratio is smaller which means that the ratio is $z$ dependent. 
The ratio is practically independent of the collision energy.
In the present calculations, 
the $z$-averaged ratio for $|\cos \theta|<$ 0.6 is about $0.2$.
The experimental error bars for the ratio (only statistical) were obtained 
with the help of the following formula:
\begin{equation}
\Delta \left(\frac{\sigma (\pi^0 \pi^0)}{\sigma (\pi^+ \pi^-)}\right) = 
\sqrt{\left( \frac{1}{\sigma (\pi^+ \pi^-)} \right)^2 
\Delta^2 \sigma (\pi^0 \pi^0) +
\left( \frac{\sigma (\pi^0 \pi^0)}{\sigma^2 (\pi^+ \pi^-)} \right)^2 
\Delta^2 \sigma (\pi^+ \pi^-)}.
\end{equation} 
The experimental data points are in between the predictions of the BL pQCD
approach and the hand-bag model which further clouds the situation.

\section{The nuclear cross section for the pion pair production}

\begin{figure}[!h]                
\includegraphics[width=0.3\textwidth]{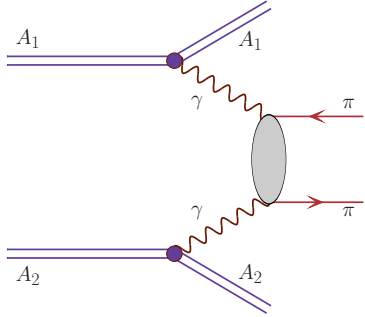}
   \caption{\label{fig:nuclear}\textsl{}
   \small The Feynman diagram illustrating the formation of the pion pair
   as a result of the peripheral nuclear collision.
}
\end{figure}
In our opinion the equivalent photon approximation 
in the impact parameter space (b-space EPA) is the best suited approach
for applications to the peripheral collisions of nuclei.
In this approach absorption effect can be taken into account easily 
by limiting impact parameter $b > R_1 + R_2 \approx 14$ fm.
This approach have been used recently in the calculation 
of the muon pairs or $\rho^0 \rho^0$ pairs. 
The details of the b-space EPA have been described in 
\cite{KS_rho, KS_muon}. Below we present a useful and compact
formula for calculating the total cross section 
for the considered process:
 \begin{eqnarray}
\sigma \left(Pb Pb \rightarrow Pb Pb \pi \pi ; W_{\gamma \gamma}\right)
= \int  {\hat \sigma}\left(\gamma\gamma\rightarrow \pi \pi; W_{\gamma \gamma}
\right) \theta \left(|{\bf b}_1-{\bf b}_2|-2R_A \right)& &
\nonumber \\
\times   N \left(\omega_1,{\bf b}_1 \right) N\left(\omega_2,{\bf
b}_2 \right)2 \pi b \, {\rm d} b \, {\rm d} \overline{b}_x \, {\rm
d} \overline{b}_y \frac{W_{\gamma \gamma}}{2} {\rm d}W_{\gamma
\gamma} {\rm d} Y & \, & ,
 \label{eq.tot_cross_section_our}
\end{eqnarray}
where the quantities $ N \left(\omega,{\bf b} \right)$ can be
interpreted as photon fluxes associated with each of the nucleus
and $\overline{b}_x, \overline{b}_y$ are auxiliary quantities which
have been introduced in \cite{KS_quark}. 
The photon flux is expressed in terms of the charge form factor.
\begin{figure}[!h]                 
\begin{minipage}[t]{0.46\textwidth}
\centering
\includegraphics[width=1\textwidth]{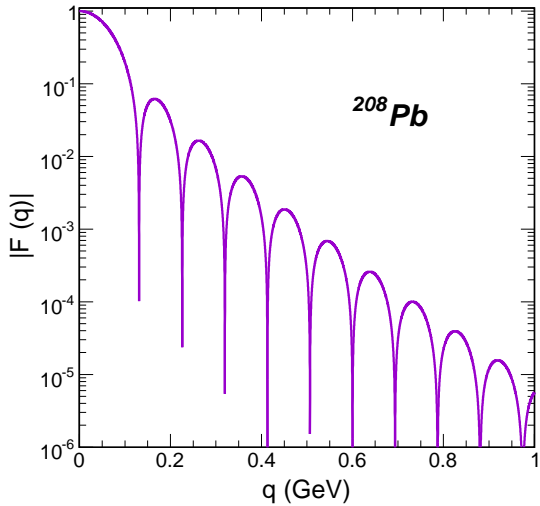}
\end{minipage}
   \caption{\label{fig:ff}\textsl{}
   \small The modulus of the charge form factor of the $^{208}$Pb
   nucleus for realistic charge distribution.
}
\end{figure}

In Fig. \ref{fig:ff} we show the modulus of the charge form factor
of the $^{208}$Pb nucleus for realistic charge distribution.
The oscillations are related to relatively sharp edge of the nucleus.

\begin{figure}[!h]                  
\begin{minipage}[t]{0.46\textwidth}
\centering
\includegraphics[width=1\textwidth]{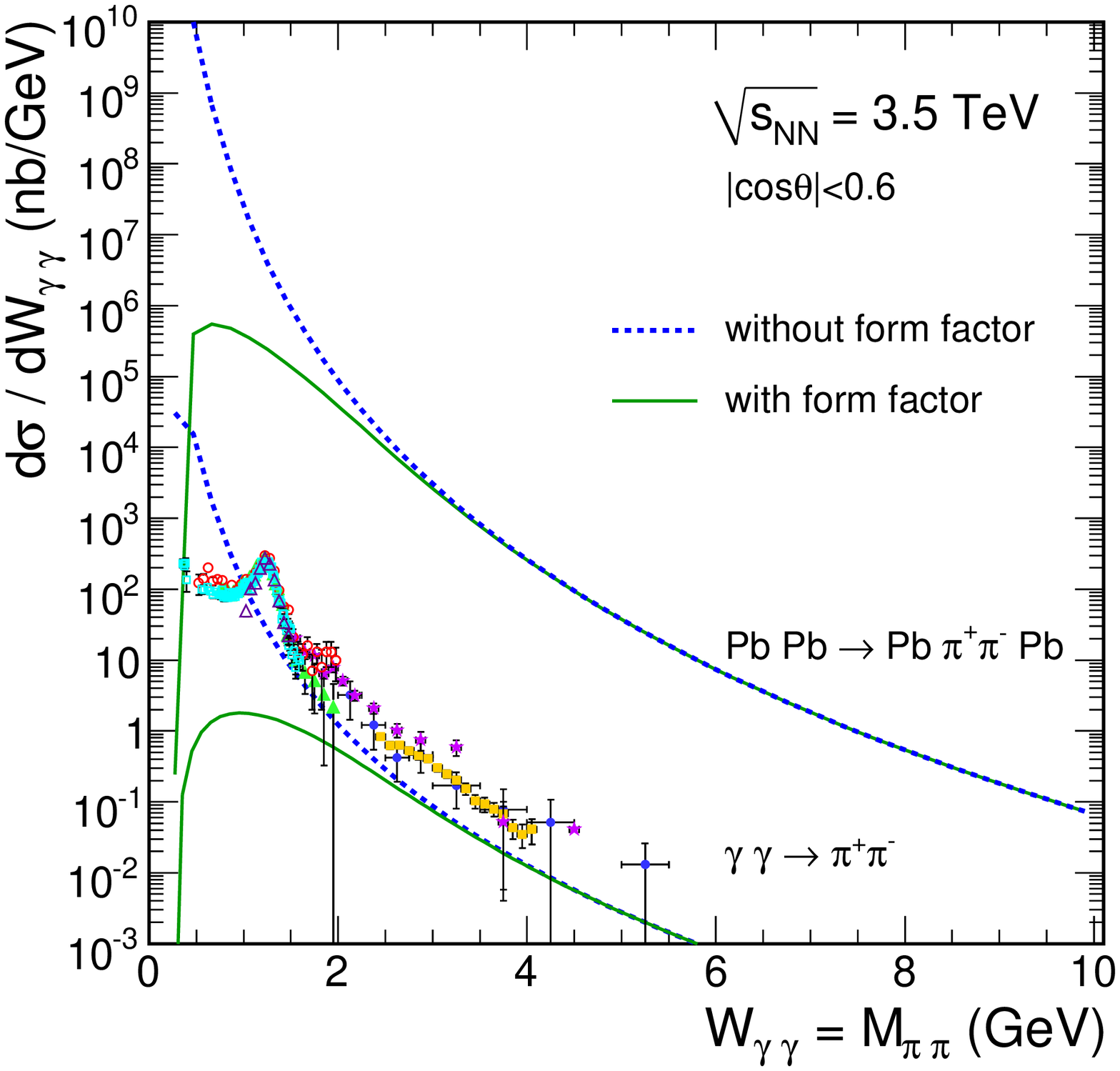}
\end{minipage}
\hspace{0.03\textwidth}
\begin{minipage}[t]{0.46\textwidth}
\centering
\includegraphics[width=1\textwidth]{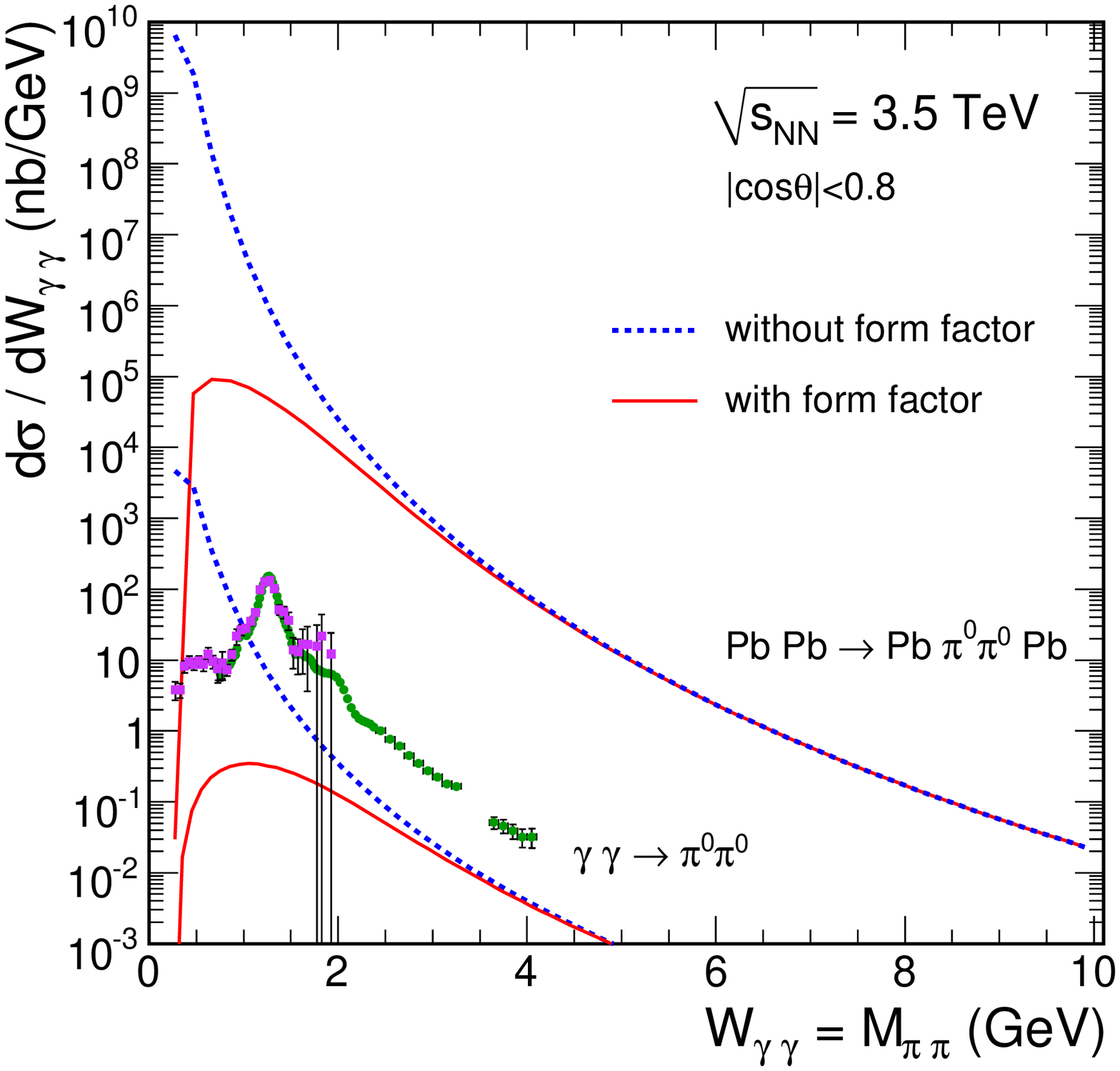}
\end{minipage}
   \caption{\label{fig:dsig_dw}\textsl{}
   \small The nuclear (upper lines) and elementary (lower lines) cross section
   as a function of photon--photon subsystem energy 
   $W_{\gamma \gamma}$ in the b-space EPA within the BL pQCD approach
   for the elementary cross section with Wu-Huang distribution amplitude. 
   The angular ranges in the figure caption
   correspond to experimental cuts.
}
\end{figure}

Let us come now to our predictions of the nuclear cross sections.
In Fig. \ref{fig:dsig_dw} we show distribution in the two-pion invariant mass
which by the energy conservation is also the photon-photon subsystem energy.
For this figure we have taken experimental limitations usually used
for the $\pi \pi$ production in $e^+ e^-$ collisions. 
In the same figure we show our results for the $\gamma \gamma$ collisions 
extracted from the $e^+ e^-$ collisions together with the corresponding 
nuclear cross sections for $\pi^+ \pi^-$ (left panel)
and $\pi^0 \pi^0$ (right panel) production.
We show the results for the standard BL pQCD approach and for the approach
proposed in Ref. \cite{HB} where an extra form factor given by
Eq. (\ref{eq.ff_pqcd}) was used to remove nonperturbative regions
of small-angle scattering described at low energy in terms of
meson exchanges. One can see that a difference occurs only at
small energies which is not the subject of the present analysis. 
Above $\sqrt{s_{NN}} >$ 3 GeV the two approaches
coincide. By comparison of the elementary and nuclear cross sections
we see a large enhancement of the order of 10$^4$ which is somewhat less
than $Z_1^2 Z_2^2$ one could expect from a naive counting.
%

\begin{figure}[!h]                  
\begin{minipage}[t]{0.46\textwidth}
\centering
\includegraphics[width=1\textwidth]{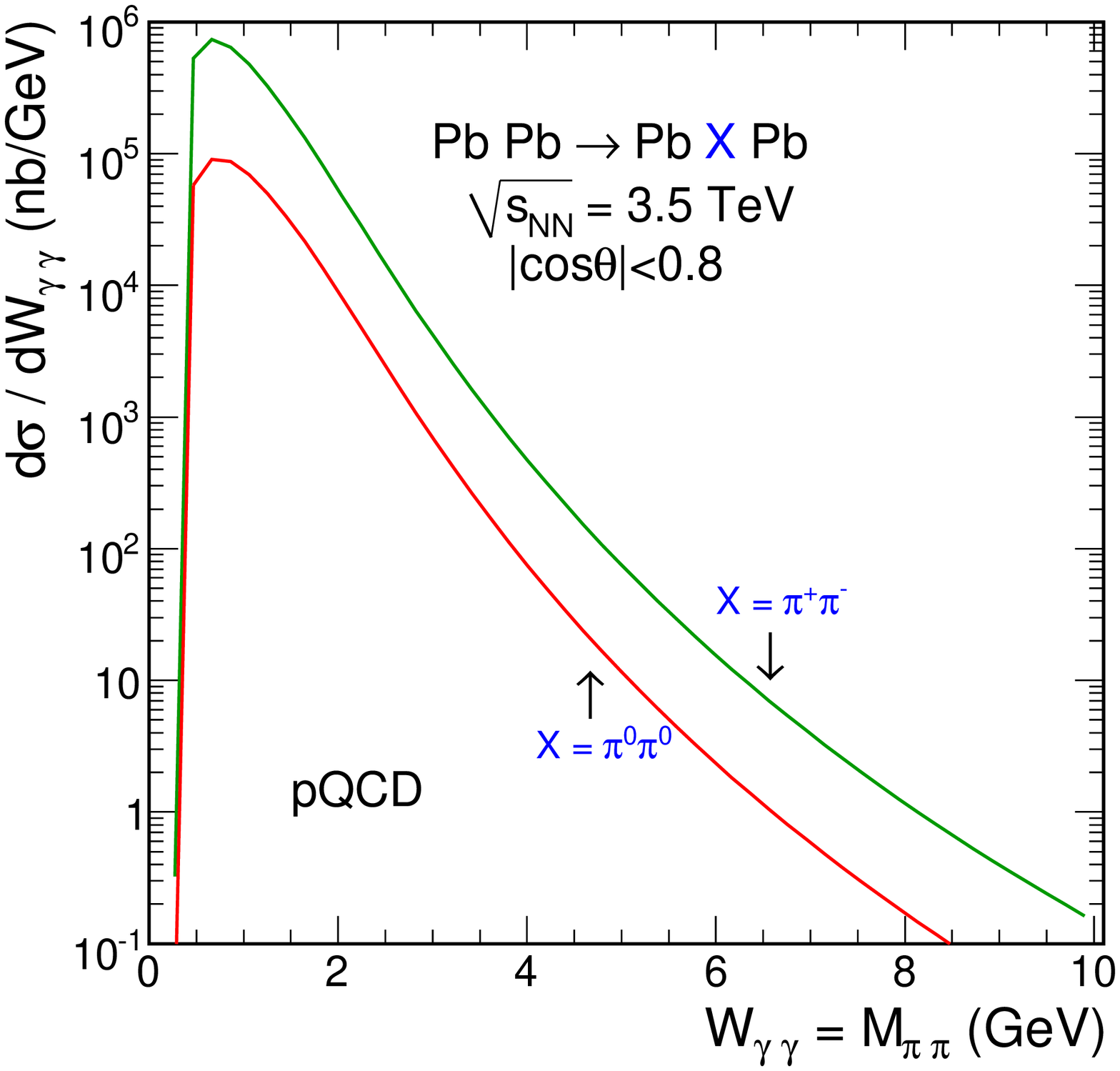}
\end{minipage}
\hspace{0.03\textwidth}
\begin{minipage}[t]{0.46\textwidth}
\centering
\includegraphics[width=1\textwidth]{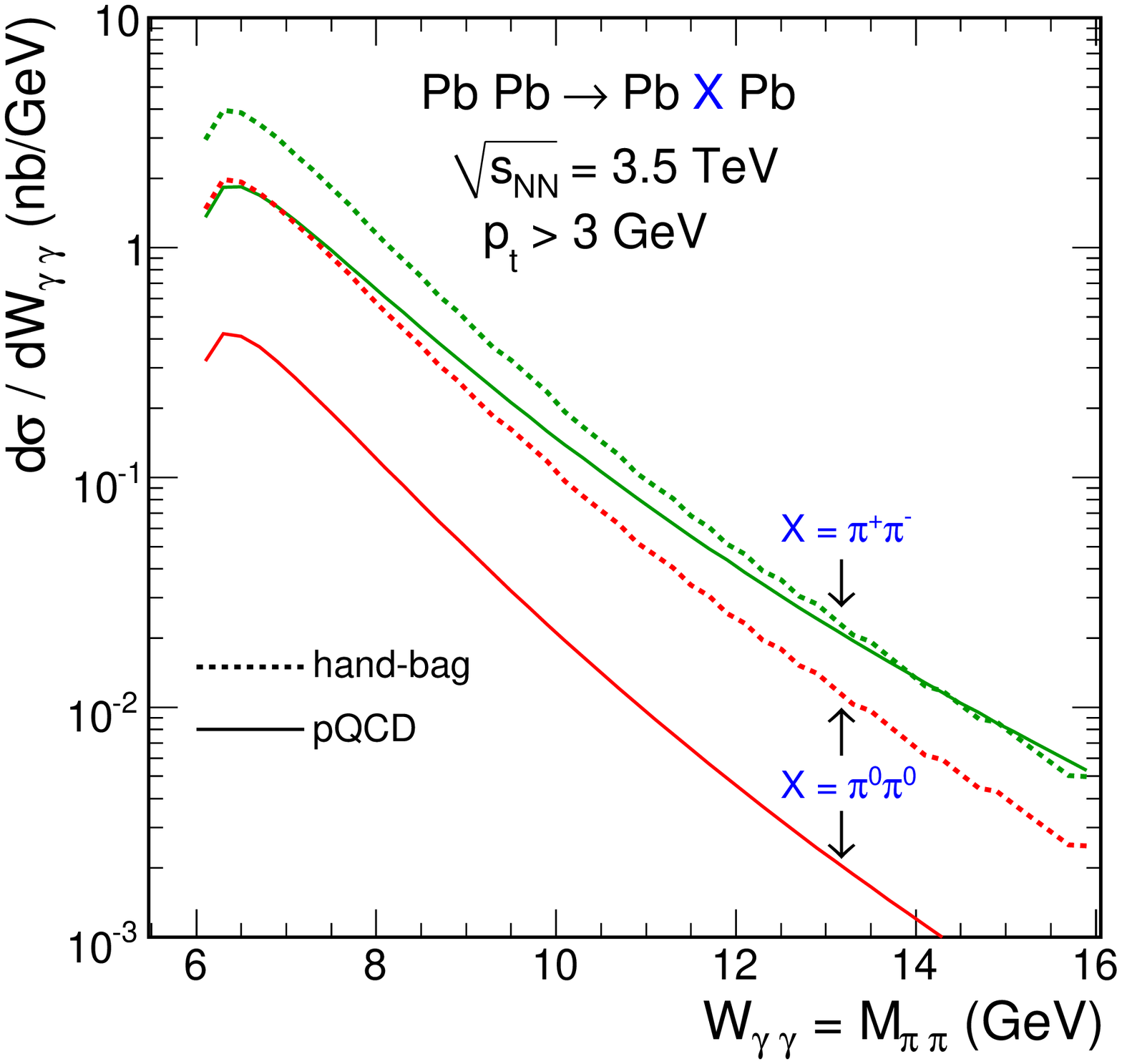}
\end{minipage}
   \caption{\label{fig:dsig_dw_z}\textsl{}
   \small The nuclear cross section as a function of 
   the $\gamma\gamma$ subsystem energy
   for the $Pb Pb \to Pb Pb \pi^+ \pi^-$ (green lines) and
   for the $Pb Pb \to Pb Pb \pi^0 \pi^0$ (red lines) reactions
   calculated for $|\cos \theta |\leq 0.8$ (left panel) 
   and with an extra cut--off on pion transverse momentum 
   $p_t >$ 3 GeV (right panel).
}
\end{figure}

In the $e^+ e^-$ collisions the cuts on $z = \cos \theta$ are usually different for
$\pi^+ \pi^-$ than for $\pi^0 \pi^0$. In the left panel of Fig. \ref{fig:dsig_dw_z}
we show the nuclear cross section for the same cut on $z$.
In the Brodsky-Lapage pQCD approach the cross section for $\pi^+ \pi^-$ production
is about order of magnitude larger than that for the $\pi^0 \pi^0$ production.
This is very different than for the hand-bag approach where the ratio
is just $\frac{1}{2}$. As already commented above one can trust the pQCD results only
for not too small energies and not too small angles or equivalently for not too
small transverse momenta of pions. In the right panel we compare results of
the Brodsky-Lepage pQCD approach (solid line) and results of 
the hand-bag approach (dashed line). Here in order to ensure validity of the 
both approaches we have imposed extra cuts on pion transverse momenta ($p_t >$ 3 GeV).
At lower energies ($W <$~14~GeV) the hand-bag cross section is bigger than 
the cross section for the Brodsky-Lepage pQCD for the $\pi^+ \pi^-$ production 
and the situation reverses at higher enrgies. 
For the $\pi^0 \pi^0$ production the hand-bag
cross section is always bigger than the BL pQCD cross section
in the shown energy range.
In this case the measured cross sections are not too big but should be measurable.

\begin{figure}[!h]                  
\begin{minipage}[t]{0.46\textwidth}
\centering
\includegraphics[width=1\textwidth]{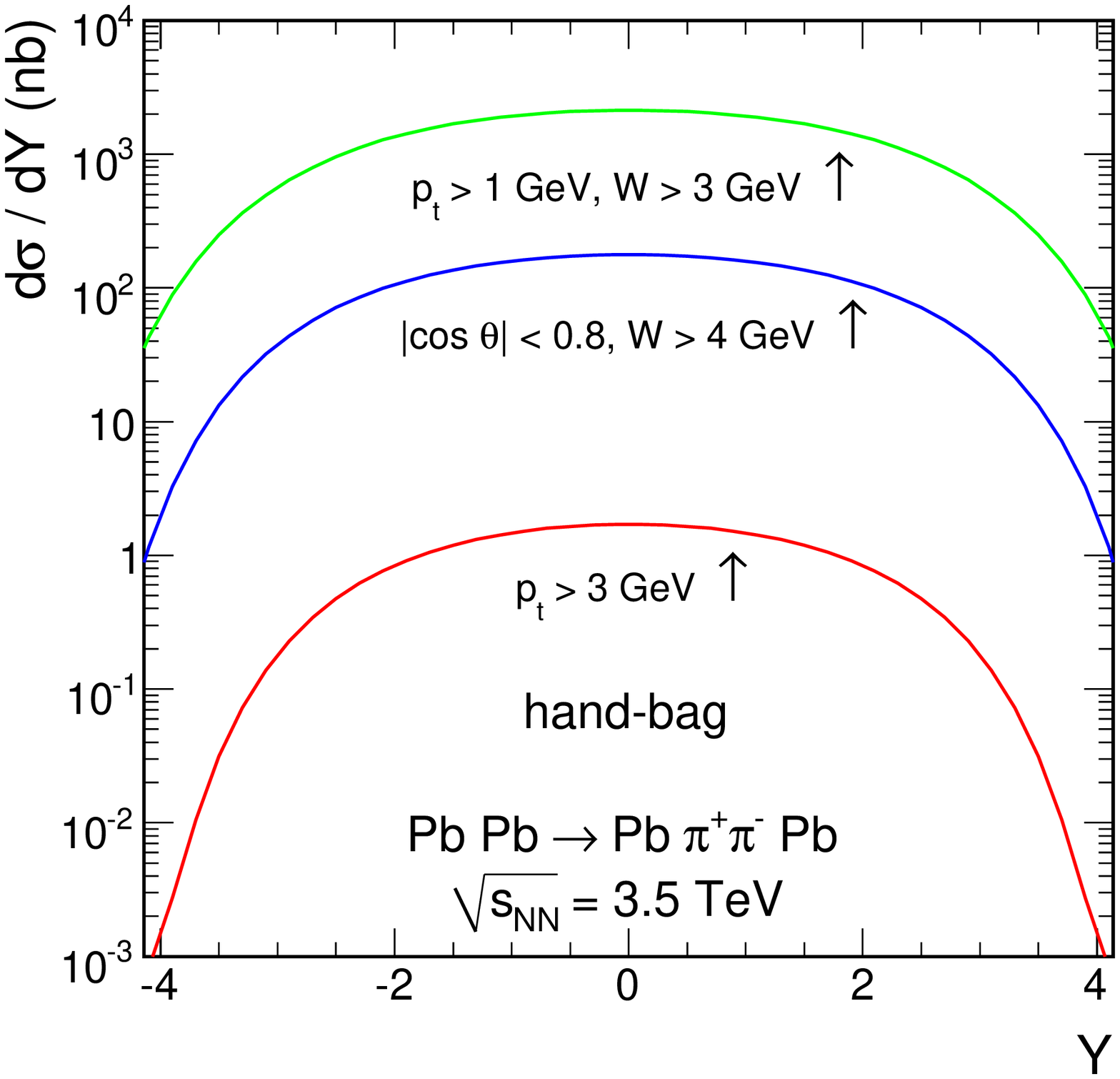}
\end{minipage}
\hspace{0.03\textwidth}
\begin{minipage}[t]{0.46\textwidth}
\centering
\includegraphics[width=1\textwidth]{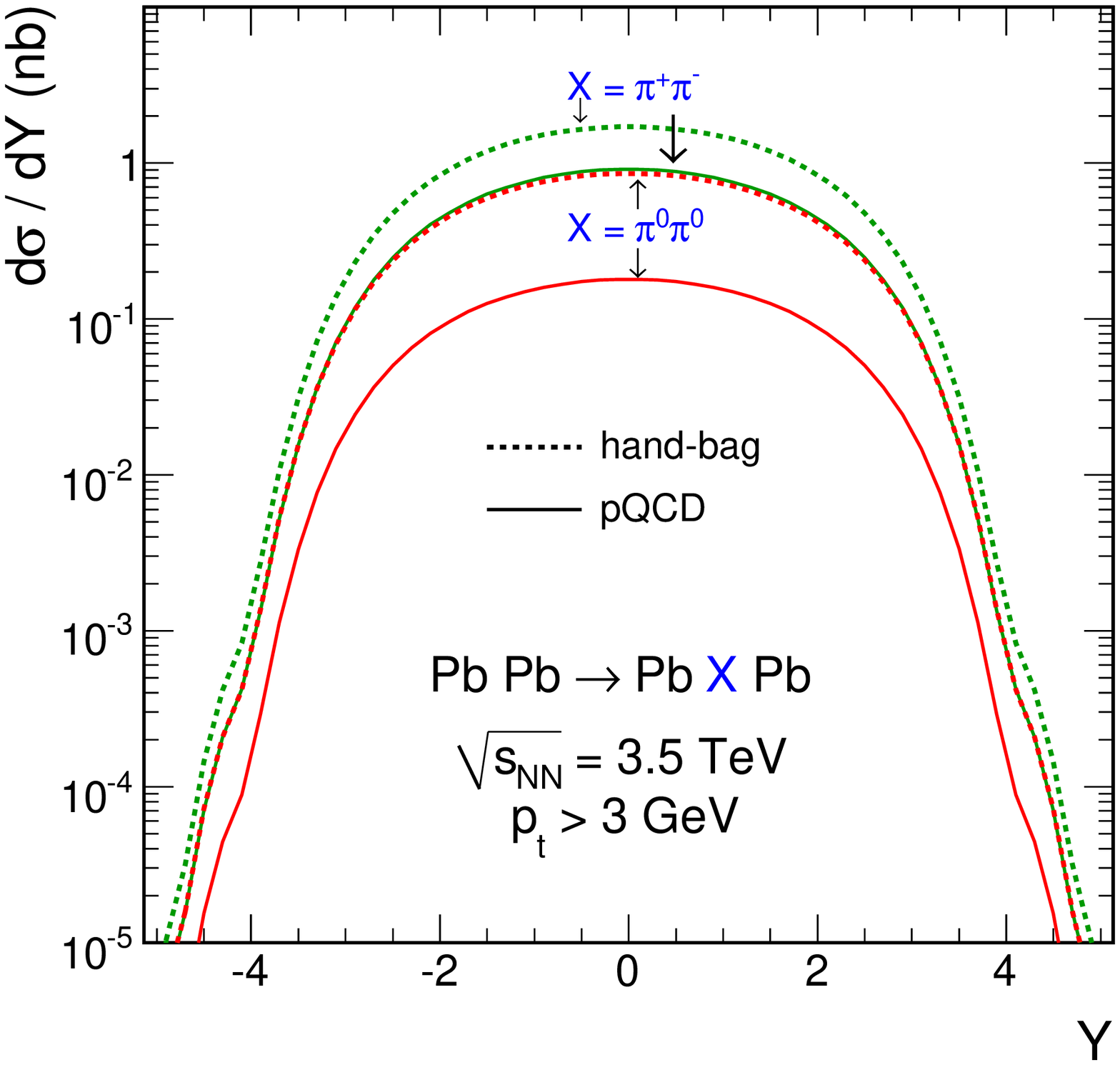}
\end{minipage}
   \caption{\label{fig:dsig_dy_z}\textsl{}
   \small The pion pair rapidity distribution. 
   The left panel shows the result for the hand-bag model 
   for different kinematic regions 
   and the right panel compares the results 
   for BL pQCD and hand-bag approaches for $p_t >$ 3 GeV,
   i.e. region not accessible so far in the $e^+ e^-$ collisions.
}
\end{figure}

As shown before the hand-bag approach better describes
the elementary cross section. Therefore the hand-bag approach
is used to estimate nuclear cross section.
In the left panel of Fig. \ref{fig:dsig_dy_z} we show
pion pair rapidity distributions for different cuts.
We hope that this figure may be a useful estimate of the
cross sections for possible future experiments.
%
In the right panel of Fig. \ref{fig:dsig_dy_z} we compare the results of 
the BL pQCD approach and of the hand-bag approach for $p_t >$ 3 GeV 
(which by kinematics is equivalent to $W_{\gamma \gamma} >$ 6 GeV). 
This is a region which was not measured so far in the $e^+ e^-$ collisions.
Nuclear experiment in this region should therefore discriminate
between the two approaches. One could measure either integrated cross section
with cuts as well as study the ratio for $\pi^0 \pi^0$ to 
$\pi^+ \pi^-$ as a function of accessible kinematical variables.

\section{Conclusions}

In the present paper we have discussed a possibility to study the
$\gamma \gamma \to \pi \pi$ processes in ultraperipheral ultrarelativistic
heavy-ion collisions.

In the present paper we have concentrated on the large two-pion invariant masses.
First, we show how different reaction mechansims describe the 
large photon-photon energy data. We have discussed 
the pQCD Brodsky-Lepage mechanism with 
the distribution amplitude used recently to describe the pion 
transition form factors measured by the BABAR collaboration. 
For comparison we have considered the soft hand-bag mechanism 
proposed by Diehl, Kroll and Vogt.
In addition we have considered also $t$ and $u$ channel $\omega$ meson exchanges.
In our opinion the situation in the measured energy range 
$\sqrt{s_{\gamma \gamma}} <$ 4 GeV is not clear.

The elementary cross sections have been used to make predictions
for the exclusive production of pionic pairs in heavy-ion collisions.
In order to concentrate on the interesting region where the pQCD may apply
we have imposed cuts on pion angles in the dipion center of mass 
and on the pion transverse momenta. In addition, this allows to get rid of 
the soft and resonance regions. In the present paper
we have presented predictions for the present LHC energy
$\sqrt{s_{NN}}$ = 3.5 TeV. 
The distributions in the two-pion invariant mass
and pion-pair rapidity have been calculated and shown.

Both the STAR collaboration at RHIC and the ALICE collaboration at LHC 
could measure the cross section for the exclusive $\pi^+ \pi^-$ 
production not only in the perturbative region. The region of 
resonances can be measured already with low statistics. 
Since the cross section for large invariant masses is smaller it
requires good statistics.
Having the absolutely normalized cross sections is very important
in this context. In general diffractive nuclear photon-pomeron 
mechanism can also contribute to the discussed region. Such a process
is naively enhanced in nuclear collisions only by the $Z^2$ factor compared 
to the $Z^4$ factor for the mechanism discussed here. 
A real comparison to future
data will require inclusion of the mechanism too. This goes, however,
beyond the scope of the present analysis and requires further development
in understanding nuclear diffractive processes. This is on our list
of the topics of interest.

\vspace{1cm}

{\bf Acknoweledgments}

We are indebted to Christoph Mayer for a discussion of the
capability of the ALICE detector
of measuring exclusive production of two charged pions
and Nikolai Achasov for the discussion of some problems related to
the $\gamma \gamma \to \pi^+ \pi^-$ reaction.
\\
This work was partially supported by the 
Polish grant N N202 078735 and N N202 236640.

\end{document}